\documentclass[%
 aip,
 graphicx
 amsmath,amssymb,
 reprint,%
]{revtex4-1}

\draft 

\usepackage{graphicx}
\usepackage{dcolumn}
\usepackage{bm}

\usepackage[utf8]{inputenc}
\usepackage[T1]{fontenc}
\usepackage{mathptmx}
\usepackage{etoolbox}
\usepackage{float}

\usepackage{ulem}
\usepackage{xcolor}
\definecolor{darkgreen}{rgb}{0.,0.5,0.}
\definecolor{lightgreen}{rgb}{0.,0.76,0.76}
\definecolor{darkblue}{rgb}{0.1,0.2,0.46}
\definecolor{lightred}{rgb}{1.0,0.3,0.3}
\definecolor{darkred}{rgb}{.8,0.1,0.0}
\definecolor{lilac}{rgb}{0.6,0.2,0.7}

\newcommand*\TAC{\ensuremath{T_{\rm AC}}}
\newcommand*\TMC{\ensuremath{T_{\rm MC}}}
\newcommand*\thetaSK{\ensuremath{\theta_{\rm sk}}}
\renewcommand{\vec}[1]{\boldsymbol{\mathbf{#1}}}

\makeatletter
\def\@email#1#2{%
 \endgroup
 \patchcmd{\titleblock@produce}
  {\frontmatter@RRAPformat}
  {\frontmatter@RRAPformat{\produce@RRAP{*#1\href{mailto:#2}{#2}}}\frontmatter@RRAPformat}
  {}{}
}%
\makeatother
\begin{document}

\preprint{AIP/123-QED}

\title[Reversal of the skyrmion topological deflection across ferrimagnetic angular momentum compensation]{Reversal of the skyrmion topological deflection across ferrimagnetic angular momentum compensation}

\author{L. Berges}
\author{R. Weil}
\author{A. Mougin}
\author{J. Sampaio}%
 \email{joao.sampaio@universite-paris-saclay.fr}
 
\affiliation{ 
Universit\'e Paris-Saclay, CNRS, Laboratoire de Physique des Solides, 91405 Orsay, France
}

\date{\today}

\begin{abstract}
   Due to their non-trivial topology, skyrmions describe deflected trajectories, which hinders their straight propagation in nanotracks and can lead to their annihilation at the track edges. This deflection is caused by a gyrotropic force proportional to the topological charge and the angular momentum density of the host film.  In this article we present clear evidence of the reversal of the topological deflection angle of skyrmions with the sign of angular momentum density. We measured the skyrmion trajectories across the angular momentum compensation temperature (\TAC{}) in GdCo thin films, a rare earth/transition metal ferrimagnetic alloy. The sample composition was used to engineer the skyrmion stability  below and above the \TAC{}. A refined comparison of their dynamical properties evidenced a reversal of the skyrmions deflection angle with the total angular momentum density. This reversal is a clear demonstration of the possibility of tuning the skyrmion deflection angle in ferrimagnetic materials and paves the way for deflection-free skyrmion devices. 
\end{abstract}

\maketitle


The discovery of efficient driving of chiral magnetic textures by current-induced spin-orbit torques \cite{Moore2008,Thiaville2012a,Manchon2019} has opened the possibility of energy-efficient and high-performance spintronic devices \cite{Fert2017b,Sampaio2013}, with applications in
digital~\cite{Zhang2015a} or 
neuromorphic~\cite{Huang2017,Zazvorka2019,Sai2017,Song2020a} computation, 
ultra-dense data-storage~\cite{Fert2013c,Brataas2012},
and signal processing \cite{Carpentieri2015,Finocchio2015}. Chiral textures are stable in magnetic thin films with a significant Dzyaloshinskii-Moriya interaction (DMI), typically induced with an adjacent heavy-metal layer (e.g. Pt/Co). Additionally, the heavy-metal layer, through the spin Hall effect, converts an applied charge current into a spin current that drives the magnetic textures by spin orbit torque (SOT). Very promising mobility of chiral magnetic domain walls (DW) has been observed \cite{Moore2008,Kim2017}, with nonetheless a saturating mobility at large current densities \cite{Thiaville2012a}.
Another archetypal chiral magnetic texture is the skyrmion, a small (down to few tens of nm) radially symmetric whirling texture. Although highly mobile \cite{Boulle2016,Hrabec2017c,Jiang2017a,Reichhardt2022}, their non-trivial topology induces a transverse deflection of their trajectory, a phenomenon known as gyrotropic deflection or skyrmion Hall effect \cite{Zang2011,Jiang2017a,Chen2017}. This reduces the velocity in the forward direction and can lead to the annihilation of the skyrmion at the edges of the hosting magnetic track, and is thus highly undesired.  

The gyrotropic deflection can be mitigated in magnetic systems with anti-parallel lattices \cite{Dohi2019}, such as antiferromagnets or ferrimagnets, where the overall angular momentum density of the double skyrmion can be suppressed.
In particular, ferrimagnetic alloys of the rare-earth/transition-metal (RETM) family, where the RE and TM moments are antiferromagnetically coupled \cite{Hansen1989,Sala2022}, are a promising example. In a previous work by our team, it was shown that skyrmions in GdCo thin films attained the high-mobility linear regime beyond pinning, and that their velocity and deflection followed the predictions of the Thiele model \cite{Berges2022}.
However,  there is still only little experimental evidence of the advantages of these systems~\cite{Woo2018b,Caretta2018b}, especially regarding the control of the gyrotropic deflection. 
 
In RETMs, The balance between the moments of different nature can be changed with alloy composition or temperature which leads to two points of interest for skyrmions. At the first one, the magnetic compensation temperature \TMC{}, the magnetization of the two sub-lattices are equal, the total magnetization ($M_s = M_{\rm TM} - M_{\rm RE}$) vanishes, and the size of the skyrmions is minimal due to the absence of dipolar fields~\cite{Caretta2018b}.
As  RE and TM have different gyromagnetic ratios ($\gamma_{\rm RE}$ and $\gamma_{\rm TM}$), the total angular momentum density ($L_s = \frac{M_{\rm TM}}{\gamma_{\rm TM}}  - \frac{M_{\rm RE}}{\gamma_{\rm RE}}$) will vanish at a different temperature, the angular compensation temperature \TAC{}. Both \TMC{}  and \TAC{} depend on composition. The reduction and reversal of the total angular momentum, which is the root cause of magnetic precession, leads to interesting dynamical properties near \TAC{}, such as e.g. the reversal of the deflection angle of chiral domain wall fingers \cite{Hirata2019} or the precessionless motion of magnetic domains walls \cite{Haltz2020}. However, the reversal of the skyrmion gyrotropic deflection at \TAC{} has not yet been demonstrated.

In this letter, we measure the velocity and deflection angle of skyrmions driven by spin-orbit torques in two Pt/GdCo/Ta films of different composition, above and below their \TAC{}. 
We show the dependence of the deflection with angular moment density, and in particular its reversal by changing sample composition or temperature. A quantitative analysis with a rigid texture model based on the Thiele equation is used to characterize the role of the material parameters on the skyrmion dynamics. 

\begin{figure}[h]
\includegraphics[width=0.9 \columnwidth]{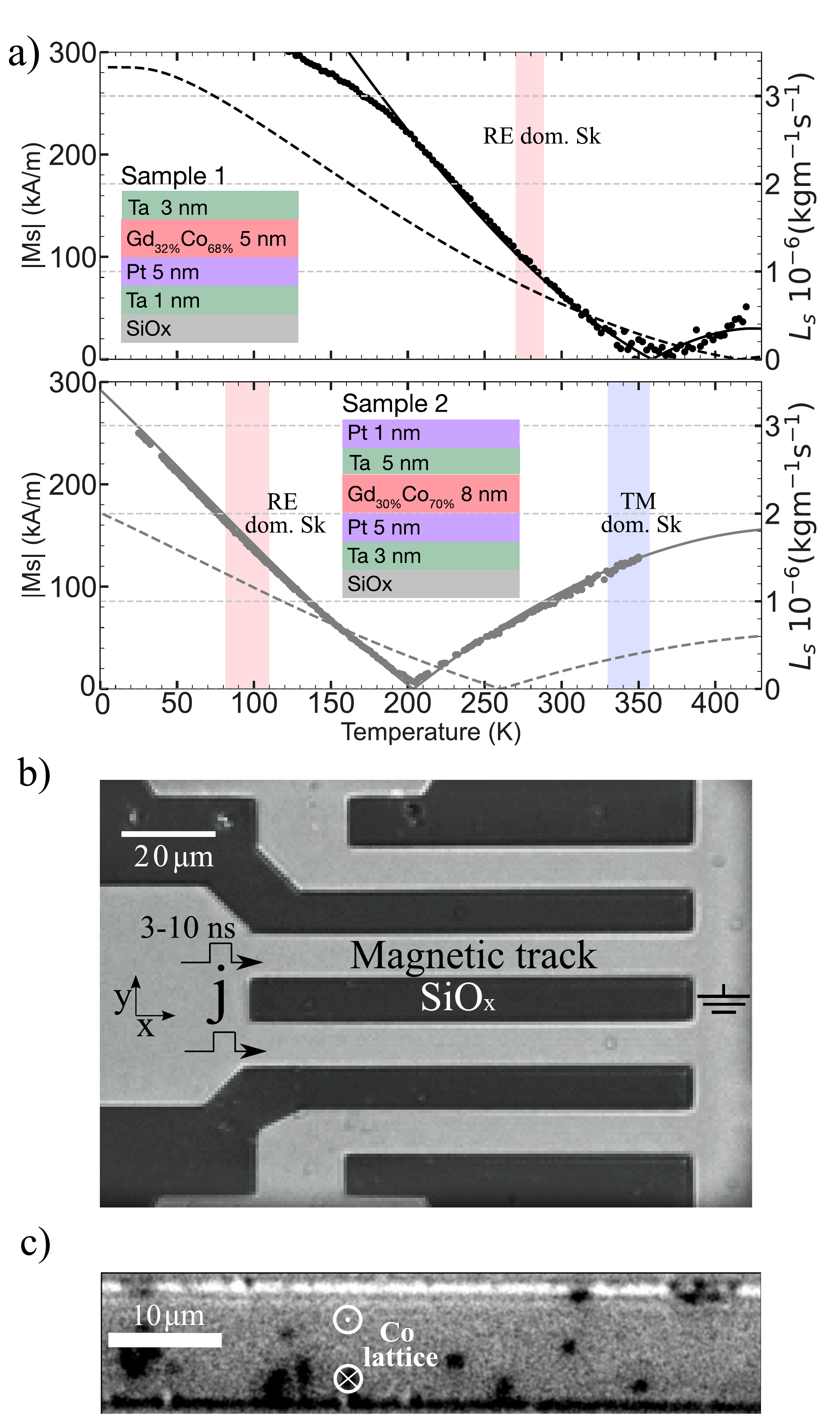}
\caption{\label{fig:Fig1} 
a) $M_s$ versus temperature for sample 1 (top panel, black points) and sample 2 (bottom panel, gray points) measured by SQUID (Superconducting Quantum Interference Device) magnetometry, and mean-field-computed curves of $M_s$ (solid line) and $L_s$ (dashed line) for both samples. Samples stacks are presented in insets and skyrmion temperature stability regions in colored bands. b) Typical magnetic device studied for sample 2. c) Examples of differential MOKE images obtained in sample 2 at 350~K. }
\end{figure}

The skyrmion dynamics were measured in two samples. Sample 1 is composed of a film of (Si/SiOx(100))/ Ta(1)/ Pt(5)/ Gd$_{0.32}$Co$_{0.68}$(5)/ Ta(3) and sample 2 of (Si/SiOx(300))/ Ta(3)/ Pt(5)/ Gd$_{0.3}$Co$_{0.7}$(8)/ Ta(5)/ Pt(1) (thicknesses in nm) as presented in the insets in Fig.~\ref{fig:Fig1}a.
The samples were patterned into 10~$\mu$m- or 20~$\mu$m-wide tracks in order to apply current pulses (Fig.\ref{fig:Fig1}b).
The magnetization as a function of temperature was measured by SQUID magnetometry on unpatterned samples and is presented in Fig.~\ref{fig:Fig1}(a). Sample 1 presents a \TMC{} around 360~K whereas sample 2 presents a \TMC{} around 200~K. Therefore, at room temperature, sample 1 is RE-dominated whereas sample 2 is TM-dominated, where RE or TM domination refers to which sublattice has the higher magnetic moment and therefore  aligns with an external magnetic field.
It is useful to use the effective ferromagnet model of ferrimagnets \cite{Wangsness1953}, which assumes a signed magnetization and angular momentum density  that are positive, by convention, when TM-dominated: $M_s=|M_{\rm Co}|-M_{\rm Gd}|$ and $L_s=|L_{\rm Co}|-L_{\rm Gd}|$. 
The exact determination of the \TAC{} is not straightforward. It was therefore deduced for both samples, using the mean field model described in ref. \cite{Berges2022}. The calculated $L_S(T)$ are shown by the dashed lines in Fig.~\ref{fig:Fig1}(a), and yield $\TAC{}= 416$~K for sample 1 and $\TAC{}= 260$~K for sample 2. These results are consistent with the empirical law described in ref.~\cite{Hirata2018} which gives \TAC{} for GdCo between 40 to 60~K above the \TMC{}.

The magnetic textures are observed in each sample as a function of temperature by magneto-optical-Kerr-effect (MOKE) microscopy. 
A typical differential MOKE image is presented in Fig.~\ref{fig:Fig1}(c).
Skyrmions are observed in the temperature ranges indicated by the color bands in Fig.~\ref{fig:Fig1}(a). In these ranges, starting from a saturated state and lowering the applied external magnetic field, skyrmions with a core of opposing  magnetization will naturally nucleate at small enough field ($-30$ to 0 mT for an initial saturation at large negative magnetic field). Skyrmions can also be nucleated by applying electrical 
pulses~\cite{Berges2022,Quessab2022}.
A typical phase diagram (versus temperature and field) of these samples is presented in a previous work \cite{Berges2022}. In the studied temperature range, sample 1 only presents one skyrmion stability range around 290~K, whereas sample 2 presents two skyrmion stability ranges, one around 90~K and a second around 350~K. 
In sample 1, the skyrmion stability range is below \TMC{} (and \TAC{}), where the film $L_S<0$, and so these are dubbed RE-dominated skyrmions. In sample 2, the skyrmions at 90~K are RE-dominated as well, while the skyrmions at 350~K are TM-dominated (above \TMC{} and \TAC{} with therefore $L_S>0$).
Note that in the MOKE images, the signal is proportional to the Co sublattice, independently of the temperature \cite{bergesphd2022}. Thus, skyrmions with a core Co moment pointing along the same direction  will appear with the same color (black for $-z$ with our experimental conditions), whether they are RE- or TM-dominated (Fig.~\ref{fig:Fig1}c).

\begin{figure*}[]
    \centering
    \includegraphics[width=0.8 \textwidth]{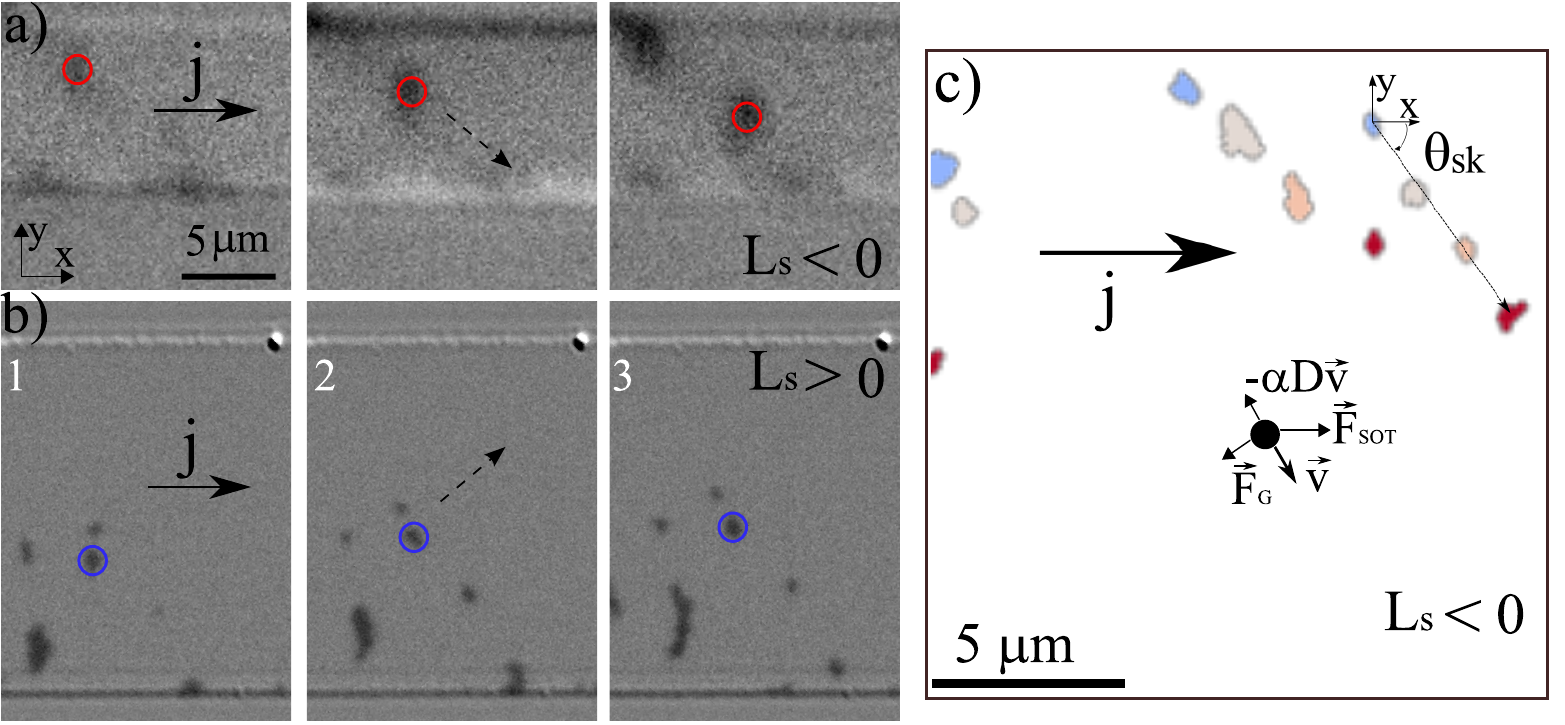}
    \caption{ 
    a)-b) Example of three successive MOKE images separated by 10-ns (6 ns) pulses showing the displacement of skyrmions in sample 2 with at a) at T = 110~K and $j=120$~GA/m$^2$  and b) T = 350~K and 150~GA/m$^2$ (field around $0$~mT.) The colored circles identify the same skyrmions in the three images. 
    As the images in a) were obtained using a cryostat and are of lower resolution, a different temperature from the one used in the dynamical studies (90~K) was used to render the skyrmions larger and more visible. 
    c) Superposition of four consecutive MOKE images, in the case of sample 1 with $L_s<0$, showing the propagation of three skyrmions for 2 and 4 images. The different forces defined in the Thiele equation acting on the skyrmions are sketched around the black dot. 
    }
    \label{fig:Fig2}
\end{figure*}

Once skyrmions are nucleated, electrical pulses of 3 to 10~ns are applied and MOKE images are acquired in order to study the skyrmions dynamics. 
The skyrmion motion is tracked over several pulses using a partially-automated process described in ref.~\cite{Berges2022}, and their velocity and deflection are calculated considering the pulse duration and the traveled distance. Typical images of skyrmions displacements are shown in Fig.~\ref{fig:Fig2}, in the case of sample 2 at low temperature and $L_s<0$ (a) and high temperature and $L_s>0$ (b). 
The average skyrmion diameter was similar for the three studied cases, 0.86$\pm$0.28~$\mu$m.
An example of the observed skyrmion dynamics in sample 1 is presented in Fig.~\ref{fig:Fig2}(c) with a superposition of successive MOKE images where the skyrmion color refers to the MOKE image number.\\

\begin{figure}[]
 \centering
    \includegraphics[width=0.9 \columnwidth]{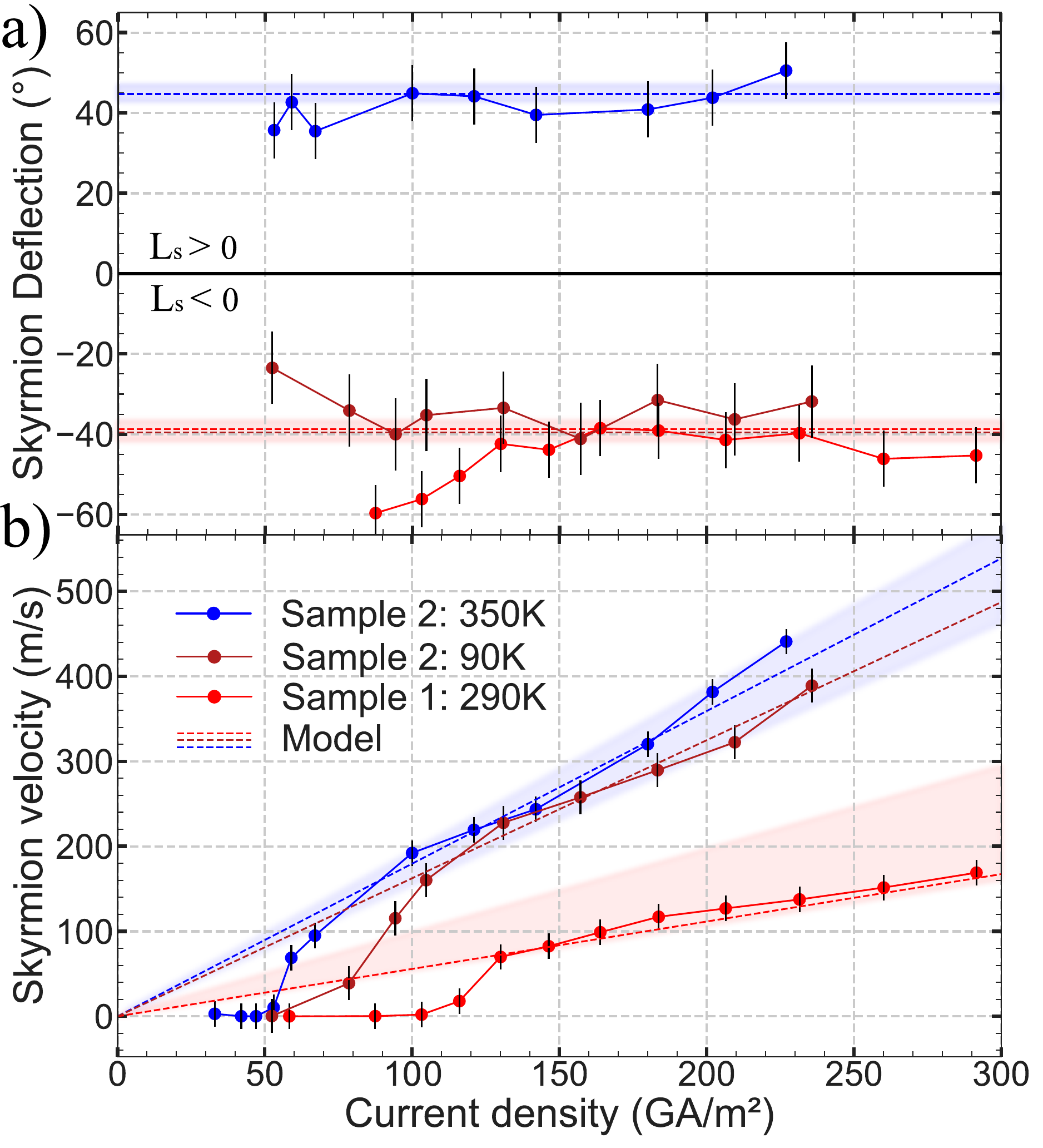}
\caption{\label{fig:Fig3} a) Averaged skyrmion deflection \thetaSK{}, and b) velocity, measured in sample 1 at 290~K and 2 at 90~K and 350~K. The error bars represent the standard deviation of the measurements. The dashed lines are obtained with the Thiele based model (with fitted $\theta_{\rm SHE}$). The color bands represent the expected error due to the experimental error of $\theta_{\rm SHE}$, as described in the main text.}
\end{figure}

The skyrmion deflection (\thetaSK{}) and velocity ($v$) versus applied current density ($j$) are presented in Fig.~\ref{fig:Fig3}(a,b) for the three cases: RE-dominated skyrmions in sample 1, and RE- and TM-dominated skyrmions in sample 2. Videos of successive displacements in both samples are shown in S.I.
In the three cases, the velocity shows a clear depinning transition above a current threshold (different for each case), and then follows a linear regime.
The mobility in the linear regime (i.e. $\Delta v/\Delta j$) is much higher in sample 2 than in sample 1. 
In sample 2, the mobility of TM-dominated skyrmions is slightly  higher than RE-dominated skyrmions. These differences in mobility will be discussed later.
The linear regime extends up to 190 m/s in sample 1 and to 450 m/s in sample 2. At highest $j$, skyrmions are nucleated by the pulse,  which hinders the tracking analysis and thus limits the maximum $j$ that can be examined.  
In the linear regime, the deflection angle \thetaSK{} is approximately constant with the current density, and its absolute value is about $40^\circ$ for the three cases. The deflection angle is clearly reversed between the TM- and RE-dominated skyrmions: it is positive for TM-dominated skyrmions (in sample 2) and negative for RE-dominated skyrmions (in both samples). 
The deflection also reverses with core polarity, i.e. with the Co moment pointing along $+z$ (which appear as white skyrmions in the MOKE images; see Supplemental Materials).
The \thetaSK{} in the pining regime is measured to be larger than in the flow regime in sample 1, whereas it is lower in sample 2. This is perhaps a bias induced by the different nucleation protocol used in these measurements. For sample 1, skyrmions were only nucleated by current pulses, mostly near one of the edges
due to the Oersted field~\cite{Berges2022}, whereas for sample 2 they were first nucleated homogeneously by magnetic field. As skyrmions can be annihilated at the edges, only the skyrmions that deviate towards the center are accounted for, which biases the measurement of the mean \thetaSK{}.\\ 

The skyrmion dynamics in the linear regime can be quantitatively analyzed using a rigid-texture formalism based on the Thiele equation~\cite{Thiele1974}. It expresses the equilibrium of all forces applied on the magnetic texture that reads in our case as: $\vec{F}_G + \vec{F}_{\rm SOT} + \alpha {\rm D} \vec{v} = \vec{0} $, where $\vec{F}_{\rm SOT}$ is the SOT force, $\vec{F}_G $ the gyrotropic force and $\alpha {\rm D}$ is, in general, a tensor describing the dissipation. This formalism can be applied to skyrmions in  double-lattice systems as presented in refs.~\cite{Berges2022,Panigrahy2022}.
These forces are depicted in Fig.~\ref{fig:Fig2} c), on a black dot representing a skyrmion in the case of $L_s<0$. The norm of the skyrmion velocity $|v|$ and its deflection \thetaSK{} can be deduced to be:
	\begin{eqnarray} \label{eq:v_theta} 
    |v|=\frac{v_0}{\sqrt{1+\rho^2}}\\
    \thetaSK{} =\arctan(\rho)
	\end{eqnarray}
In the limit of skyrmions larger than  the domain wall width parameter $\Delta$, the parameters $v_{0}$ and $\rho$ are:
	\begin{eqnarray} \label{eq:ThieleSolEffPars} 
	v_{0}  &\approx&  -\frac{\pi \Delta}{2 L_\alpha} \frac{\hbar j \theta_{\rm SHE} }{ 2 e  t}\\
	\rho &\approx& \frac{\Delta }{2\pi R} \frac{L_S}{L_\alpha} n
	\end{eqnarray}
where 
$\hbar$ is the Planck constant, $e$ the fundamental charge, $t$ the magnetic film thickness,
$\theta_{\rm SHE}$ is the effective SHE angle in the Pt layer, $L_{\alpha}=L_S \alpha$ the energy dissipation rate,  
$n= p_{\rm Co} 4\pi = \pm 4\pi$  the topological charge of the skyrmion, 
$R$ its radius, and
$p_{\rm Co}=\pm1$ is the orientation along $z$ of the core Co moment. 
Because $L_{\alpha}$ is always positive, the sign of the deflection is given by the sign of the product of $L_s$ (positive for $T>\TAC{}$) and $p_{Co}$. This sign is presented in Table~\ref{tab:tab_signs1} as a function of temperature for $p_{Co} =-1$, which is the case shown here (black skyrmions).\\

\begin{table}
\begin{ruledtabular}
     \begin{tabular}{cccc}
    & $T < \TMC{}$ & $\TMC{} < T < \TAC{}$& $T>\TAC{}$\\ \hline
    $M_{s}$ & $-$ & + & + \\ 
    $L_{s}$ & $-$ & $-$ & + \\ 
     core magnetic moment &  $+z$ $\odot$& $-z$ $\otimes$& $-z$ $\otimes$\\ 
     core cobalt moment &  $-z$ $\otimes$& $-z$ $\otimes$& $-z$ $\otimes$\\ 
     $p_{Co}$ & $-1$ & $-1$ & $-1$ \\ 
     $F_{G} \cdot y$ & $-$ & $-$ & $+$\\
     $F_{SOT}$ & $+x$ & $+x$ & $+x$\\
    \end{tabular}
\end{ruledtabular}
\caption{\label{tab:tab_signs1} 
Signs versus temperature of the material parameters, of the skyrmion core configuration parameters and of the expected forces acting on a skyrmion with negative polarity (black in the MOKE images) driven by a positive ($+x$) current.}
\end{table}

\begin{table}[]
    \centering
  \begin{ruledtabular}
  \begin{tabular}{lll}
         Parameter & Sample 1 (290 K) & Sample 2 (350 K) \\ \hline
         $\gamma$/$2\pi$ [GHz/T] \textsuperscript{a} & 18.3 & 40.8 \\
         $\alpha$ \textsuperscript{a} & 0.15&  0.175 \\
         $|M_S|$ [kA/m] \textsuperscript{b} & 78 & 125  \\
         $|L_S|$ [kg/(ms)] \textsuperscript{b}& $6.8 \times 10^{-7}$ & $4.9\times10^{-7}$  \\
         $|L_\alpha|$ [kg/(ms)] \textsuperscript{b}& $1.02 \times 10^{-7}$ & $8.5\times10^{-8}$  \\
         $\mu_0 H_k$ [mT] \textsuperscript{c} & 200 & 60 \\
         $K_u$ [kJ/m$^3$] \textsuperscript{c} & 11.5 & 13.6 \\
         $D$ [~mJ/m$^2$] \textsuperscript{a} & $-0.22$ & $-0.14$ \\
         $A$ [pJ/m] \textsuperscript{a} &  4.6 & (4.6) \textsuperscript{d}\\
         $\theta_{\rm SHE}$ \textsuperscript{e} & 0.04 & 0.09 \\ 
         $2R$ [$\mu$m] \textsuperscript{f}  & $0.85 \pm 0.28$ & $0.86 \pm 0.28$ \\
    \end{tabular}
    \end{ruledtabular}
    \caption{Measured parameters used in the model. 
    \textsuperscript{a}$\gamma$, $\alpha$, the exchange stiffness $A$ and the DMI strength $D$ (not used in the model) were determined with Brillouin light scattering (BLS) at 290~K (sample 1) or 350~K (sample 2). 
    \textsuperscript{b}$M_s(T)$ (Fig.~\ref{fig:Fig1}a) was used to determine $L_S=M_S/\gamma$ and $L_\alpha = L_S \alpha$. 
    \textsuperscript{c}$H_k$ was measured from hysteresis cycles and BLS, from which $K_u$ was deduced.
    \textsuperscript{d}$A$ was measured only on sample 1 and assumed to be the same in sample 2.
    \textsuperscript{e}$\theta_{\rm SHE}$ was determined by transport measurements using the double-harmonic technique~\cite{Berges2022,Hayashi2014} (see Supplementary Materials).
    \textsuperscript{f}The shown variation is the standard deviation of the observed radius and not the error of the average value.} 
    \label{tab:parameters}
\end{table}

The parameters needed for the model were measured on both samples (see Table~\ref{tab:parameters}).  
$H_k(T)$ was obtained by analyzing hysteresis loops at various temperatures, which yielded a value for $K_u$ ($K_u=\mu_0 H_k M_s/2 - \mu_0 M_s^2/2$) with negligible thermal variation.
The domain wall width parameter was calculated using $\Delta = \sqrt{A/K_{\rm eff}}$, where $K_{\rm eff}= \mu_0 H_k M_s /2$ is the effective anisotropy.
For Sample 2 at 90~K ($M_S=135$~kA/m), the thermal variation of $K_u$ and $A$ was neglected (as it is smaller than the precision of the other parameters) and the values at 350~K were used; $L_S=$~12.1 $\times 10^{-7}$~kg/(ms) and $L_\alpha=$10.7 $\times 10^{-8}$~kg/(ms) were deduced using a mean-field model as described in ref.~\cite{Berges2022}, assuming constant sub-lattice Gilbert damping parameters ($L_{\alpha}=\alpha_{Co}|L_s^{Co}(T)|+\alpha_{Gd}|L_s^{Gd}(T)|$). 
 The skyrmion diameter was taken from the average observed diameter, which is very similar for the three studied cases 
 \footnote{ The skyrmions show a large dispersion of diameter (Table~\ref{tab:parameters}), which we estimate to lead to a $\pm 14$\% dispersion of velocity and to a $\pm10$º dispersion of the deflection angle. However, as the values of velocity and deflection were averaged over many skyrmions and many displacements, the error due to the size dispersion is drastically reduced and is neglected.
 }.
 
These measured parameters allow to constrain the model and obtain curves for the velocity and deflection angle (dashed lines in Fig.~\ref{fig:Fig3}). A constant deflection is predicted, and its value is obtained with no fitting parameters.   
The velocity is predicted to be linear with $j$, and its slope is obtained with a single fitting parameter, $\theta_{\rm SHE}$.
The fitted values ($\theta_{\rm SHE}=$ 0.03 for sample 1 and 0.09 for sample 2) are consistent with the precision of the measured $\theta_{\rm SHE}$ (see Table~\ref{tab:parameters} and Suppl. Mat.).
The model prediction range, calculated with the estimated error  $\theta_{\rm SHE}$, is shown in the figure as a color band 
\footnote{
We estimate that the main sources of error of $\rho$ and $v$ is the spin Hall efficiency ($\theta_{\rm SHE}$). which was estimated to be 25\% (sample 1) and 12\% (sample 2); see supplemental materials.}.
Above the depinning threshold, where the model is expected to be valid, it both reproduces the qualitative behavior (constant deflection and linear velocity) and agrees quantitatively with the experimental data, within the estimated error margin. In particular, 
the sign of the deflection angle observed in the experiments agrees with Eq~(\ref{eq:ThieleSolEffPars}b) taking into account the $L_S$ of the film ($L_S<0$ for RE-dominated skyrmions and $L_S>0$ for TM-dominated skyrmions). 

The skyrmion mobility, given by the slope of the velocity versus $j$ (Fig.~\ref{fig:Fig3}(b)), is much higher in sample 2 than in sample 1  (1.80 at 350~K vs 0.6~m$\cdot$s$^{-1}$/GA$\cdot$m$^{-2}$, respectively). 
This difference in mobility cannot be ascribed to a difference in skyrmion diameter (see eq.~\ref{eq:v_theta}), as the two conditions present very similar average sizes (Table~\ref{tab:parameters}).
This large difference has multiple origins. First, the $L_\alpha$ of sample 2 is lower by 20\%.
The second major cause is the difference of the film stacks, in particular the thickness of the Ta capping layer. The measured $\theta_{\rm SHE}$ is more than twice  higher in sample 2 than in sample 1 (Table~\ref{tab:parameters}). This can be expected to be due a better passivation of the Ta layer in sample 2 which can therefore contribute more to the SOT  than the thinner (3 nm) Ta cap of the sample 1 which is probably fully oxidized. 

Finally, comparing the skyrmion velocity curves for the two conditions in sample 2 (at 90 and 350~K), it can be seen that both the depinning current and the mobility in the linear regime are significantly different.
The depinning current is higher at 90~K, which can be attributed by the thermal nature of the depinning process~\cite{Litzius2020}.
The difference in mobility is not due to a difference in skyrmion diameter (which again is very similar in all three studied conditions). 
It can be expected that several magnetic parameters vary between 90 and 350~K,  but the experimental mobility can be understood by considering only the variation of $L_{\alpha}$ ($\frac{L_{\alpha} {\rm (90 ~K)}}{L_{\alpha} {\rm (350~K)}} \approx $ 1.25).
This result and the Thiele model suggest that $L_\alpha$ is a more pertinent parameter than $\alpha$ to characterize the role of dissipation in the skyrmion mobility. Interestingly, $L_\alpha$ can  be more easily optimized than $\alpha$ to increase mobility, by increasing the sample temperature (as was the case here) or by decreasing the material's Curie temperature (all other parameters remaining equal).
A recent work~\cite{Litzius2020} on skyrmions measured at relatively high temperature also seems to point toward such an effect which seems to be an interesting path to increase skyrmion mobility.\\

In conclusion, we observed the propagation of skyrmions in the flow regime, i.e., beyond the effects of pinning in two GdCo samples, below and above the angular compensation temperature. The observed mobilities were very large, with a velocity up to 450~m/s.
The skyrmion dynamics was studied in three cases, two in RE-dominated films and one in a TM-dominated film. 
The deflection angle was constant with driving current and 
its sign was opposite between RE- and TM-dominated cases, both when comparing two samples of different composition and when comparing two temperatures (above and below \TAC{}) in the same sample. This confirms the modulation of deflection angle \thetaSK{} with $L_S$.

These experiments demonstrate the effects of the angular momentum density $L_S$ of the host material on the deflection of skyrmions. They show that \thetaSK{} can be reversed in GdCo ferrimagnetic thin films across their angular compensations, either by changing the alloy stoichiometry or simply its temperature. 
In particular, the reversal of sign of \thetaSK{} across compensation strongly supports that \thetaSK{} should be zero at angular moment compensation. The engineering of magnetic parameters that was done to produce the two presented skyrmion-hosting samples could be repeated rather straightforwardly to engineer a film with stable skyrmions at \TAC{} with no deflection.

\begin{acknowledgments}

The authors thank Stanislas Rohart for fruitful discussions, and Andr\'e Thiaville for the study of the sample properties by BLS.
This work was supported by a public grant overseen by the French National Research Agency (ANR) as part of the \textit{“Investissements d’Avenir”} program (Labex NanoSaclay, reference: ANR-10-LABX-0035, project SPICY). Magnetometry and Anomalous Hall effect measurements were performed at the LPS Physical Measurements Platform.

\end{acknowledgments}

\section*{Supplementary Material}
See supplementary material for videos of successive MOKE images showing the skyrmion motion for the three temperature regions discussed in the text. Motion of skyrmions of opposite polarity (i.e., $p_{\rm Co}=+1$; white in the MOKE images) is also shown for sample 1.
A note on the analysis of the images and the selection of relevant textures is also included. 
Experimental data of $\theta_{SHE}$ are also included for both samples at several temperatures. 

\section*{Data Availability Statement}

The data that support the findings of this study are available from the corresponding author upon reasonable request.


%

\end{document}